\documentclass[a4paper]{jpconf}
\usepackage{graphicx}
\usepackage{amsmath}
\usepackage{float}
\usepackage{amsfonts}
\usepackage{latexsym}
\usepackage{yfonts}
\usepackage{amssymb}
\usepackage{upgreek}
\usepackage{bm}
\usepackage{epsfig}
\usepackage[latin1]{inputenc}
\usepackage{rotating}
\usepackage{hyperref}

\newcommand\be{\begin{equation}}
\newcommand\ba{\begin{eqnarray}}
\newcommand\ee{\end{equation}}
\newcommand\ea{\end{eqnarray}}

\newcommand{\APO}{{\mbox{\tiny apo}}}
\newcommand{\BL}{{\mbox{\tiny BL}}}

\newcommand{\HAR}{{\mbox{\tiny H}}}

\newcommand{\INC}{{\mbox{\tiny inc}}}
\newcommand{\KERR}{{\mbox{\tiny K}}}

\newcommand{\MBH}{{{\bullet}}}
\newcommand{\met}{\mbox{g}}
\newcommand{\MIN}{{\mbox{\tiny min}}}
\newcommand{\ORB}{{\mbox{\tiny Orbital}}}
\newcommand{\PERI}{{\mbox{\tiny peri}}}
\newcommand{\PMM}{{\mbox{\tiny PM}}}
\newcommand{\REG}{{\mbox{\tiny R}}}
\newcommand{\RR}{{\mbox{\tiny RR}}}
\newcommand{\SCO}{{\star}}
\newcommand{\SF}{{\mbox{\tiny SF}}}
\newcommand{\STF}{{\mbox{\tiny STF}}}

\newcommand{\TT}{{\mbox{\tiny TT}}}

\begin{document}
\title{Approximate Waveforms for Extreme-Mass-Ratio Inspirals: The Chimera Scheme}

\author{Carlos F.~Sopuerta$^{1}$ and Nicol\'as Yunes$^{2,3,4}$}
\address{$^{1}$Institut de Ci\`encies de l'Espai (CSIC-IEEC), 
Facultat de Ci\`encies, Campus UAB, Torre C5 parells, 
Bellaterra, 08193 Barcelona, Spain.\\
${}^{2}$Department of Physics, Montana State University, Bozeman, MT 59717, USA.\\
${}^{3}$Department of Physics and MIT Kavli Institute, 77 Massachusetts Avenue, Cambridge, MA 02139, USA.\\
${}^{4}$Princeton University, Physics Department, Princeton, NJ 08544, USA.}

\ead{sopuerta@ieec.uab.es,nyunes@physics.montana.edu}

%\date{\today}

%%%%%%%%%%%%%%%%%%%%%%%%%%%%%%%%%%%%%%%%%%%%%%%%%
\begin{abstract} 
We describe a new kludge scheme to model the dynamics of generic extreme-mass-ratio inspirals 
(EMRIs; stellar compact objects spiraling into a spinning supermassive black hole) and their gravitational-wave
emission.  
The {\em Chimera} scheme is a hybrid method that combines tools from different approximation techniques in General Relativity: 
(i) A multipolar, post-Minkowskian expansion for the far-zone metric perturbation (the gravitational waveforms) 
and for the local prescription of the self-force; 
(ii) a post-Newtonian expansion for the computation of the multipole moments in terms of the trajectories; and 
(iii) a BH perturbation theory expansion when treating the trajectories as a sequence 
of self-adjusting Kerr geodesics.
The EMRI trajectory is made out of Kerr geodesic fragments joined via the method of osculating
elements as dictated by the multipolar post-Minkowskian radiation-reaction prescription.
We implemented the proper coordinate mapping between Boyer-Lindquist
coordinates, associated with the Kerr geodesics, and harmonic coordinates, associated with the
multipolar post-Minkowskian decomposition.
The Chimera scheme is thus a combination of approximations that can be used to model generic inspirals of
systems with extreme to intermediate mass ratios, and hence, it can provide valuable
information for future space-based gravitational-wave observatories, like LISA, and even for advanced ground detectors.  
The local character in time of our multipolar post-Minkowskian self-force makes this scheme 
amenable to study the possible appearance of transient resonances in generic inspirals.

\end{abstract}

\section{Introduction}
\label{intro}

The modeling of gravitational wave (GW) sources is key for the success of GW astronomy.
Detectors require precise theoretical waveforms to extract and characterize signals buried in the noise.
One of the main GW sources for space-based observatories, like the {\em Laser Interferometer Space Antenna}
(LISA)~\cite{Danzmann:2003tv,Prince:2003aa}, that requires accurate templates for detection and analysis 
are extreme-mass-ratio inspirals (EMRIs)~\cite{AmaroSeoane:2007aw}. These events 
consist of a small compact object (SCO), such as a stellar-mass black hole (BH) or a neutron 
star, spiraling in a generic orbit into a spinning, (super)massive black hole (MBH), in the regime 
where GW emission is relevant.  In such inspirals, the SCO spends up to millions of cycles in close orbits around 
the MBH, possibly with large pericenter velocities and eccentricities, sampling the strong gravitational field 
of the MBH. 

The relevant EMRI phase for data analysis is the inspiral, during which the number of cycles accumulated 
scales with the inverse of the mass ratio $q=m_{\SCO}/M_{\MBH}$ [${\cal{O}}(10^{-4})$-${\cal{O}}(10^{-6})$]. 
During EMRI plunge and merger, the accumulated number of cycles scales with the MBH mass ($M_{\MBH}$) only.  
Moreover, the signal-to-noise ratio (SNR) is in the range ${\cal{O}}(10)$-${\cal{O}}(10^{2})$ for EMRIs at realistic distances, 
and it scales linearly with the total number of cycles. Therefore, the contribution of the merger/plunge phase to the SNR
is a fraction ${\cal{O}}(m_{\SCO}/M_{\MBH})$ relative to the total inspiral contribution. 
In addition, the EMRI ringdown phase is not detectable, as the SCO barely perturbs the MBH
geometry as it plunges and crosses the MBH's event horizon. 

As a consequence of the large number of GW cycles contained in EMRIs, they can provide a detailed map of the
MBH geometry, allowing us to extract the EMRI physical parameters with high precision~\cite{Barack:2003fp}.  
This information is crucial to test the Kerr 
hypothesis~\cite{Collins:2004ex,Glampedakis:2005cf,Barack:2006pq,Vigeland:2009pr,Hughes:2010xf,Sopuerta:2010zy} 
and constrain modified gravity theories (see, e.g.~\cite{Hughes:2006pm,Schutz:2009zz,Sopuerta:2010zy,Babak:2010ej}).  
EMRI observations will also have an important impact on astrophysics (for a review see~\cite{AmaroSeoane:2007aw}).

The extreme mass ratios involved in the problem also lead to the appearance of two different spatial and
temporal scales.  The two different spatial scales are due to the different MBH and SCO sizes, $m_{\SCO}/M_{\MBH}\ll 1$. 
The two different temporal scales are the orbital one(s) and the one associated
with {\em radiation reaction}, $T^{}_{\ORB}/T^{}_{\RR} \sim m_{\SCO}/M_{\MBH}\ll 1$. Due to this, a fully relativistic 
numerical evolution of EMRIs is currently unfeasible.  

A framework to model EMRIs that exploits the extreme mass ratios involved is BH perturbation theory, where one treats 
the SCO as a small perturbation of the MBH background geometry.   In this context, the inspiral is modeled as due to  
a local {\em self-force}.  This force is composed of the metric perturbations 
generated by the SCO, regularized by eliminating divergences due to the SCO's point particle description. The SCO's motion is then
governed by the MiSaTaQuWa equation of motion, derived in~\cite{Mino:1997nk,Quinn:1997am}.   
This equation is the foundation of a self-consistent scheme to describe EMRIs in an \emph{adiabatic} way, by coupling this equation
of motion to the partial differential equations that describe the perturbations produced by the SCO. 

At present, the gravitational self-force has been computed for the case of a non-rotating MBH using 
time-domain techniques~\cite{Barack:2009ey,Barack:2010tm}. Progress is currently being made toward the calculation 
of more astrophysically relevant (spinning) EMRIs~\cite{Shah:2010bi}.  
Given the number of cycles contained produced by an EMRI in one year and 
the present complexity of self-force calculations, we cannot expect to generate complete
waveform template banks by means of full self-force calculations.  
Instead, it has been proposed that the goal of these studies
should be to understand all the details of the structure of the self-force, so that one can formulate
efficient and precise algorithms (phenomenological models) to create the waveforms needed for LISA data analysis, perhaps
complementing existent approximation schemes.

In this paper, we summarize the results of~\cite{Sopuerta:2011te} where a new {\emph{approximate}} scheme was proposed
to model EMRIs: the Chimera framework. This scheme can be thought of as a new {\em kludge} model that combines 
different approximation methods, including BH perturbation theory, the multipolar post-Minkowskian 
formalism of Blanchet and Damour~\cite{Blanchet:1984wm} and post-Newtonian theory. The Chimera scheme could be used 
for qualitative descoping studies, i.e. to determine the accuracy to which parameters could be extracted given an eLISA EMRI detection
as a function of detector properties. The Chimera scheme could also be used to study how
certain resonances, found by Flanagan and Hinderer~\cite{Flanagan:2010cd}, affect such parameter estimation and detection. 
Chimera waveforms, however, would require improvements before they can be used 
as realistic templates in data analysis. The Chimera scheme is indeed amenable to such improvements,
which will be the focus of future work. 

%%%%%%%%%%%%%%%%%%%%%%%%%%%%%%%%%%%%%%
\section{Ingredients of the Chimera Scheme}\label{chimera}

%------------------------------
\subsection{MBH Geometry and Geodesic Motion}\label{MBHgeometryandgeodesics}

The geometry of the MBH is described by the Kerr metric.
In Boyer-Lindquist (BL) coordinates, $(x^{\mu}_{\BL}) = (t,r,\theta,\phi)$, 
the line element is given by
\ba
ds^2 = -dt^2 + \frac{\rho^2}{\Delta}dr^2 + \rho^2d\theta^2 + (r^2+a^2)\sin^2\theta\, d\phi^2 
 +  \frac{2M_{\MBH}r}{\rho^2}(dt - a\sin^2\theta\, d\phi)^2 \,,
\ea
where $\rho^{2} = r^{2} + a^{2} \cos^{2}{\theta}$ and $\Delta = r^{2}-2M_{\MBH}r+a^{2} = 
r^{2} f + a^{2}$, with $f = 1 - 2 M_{\MBH}/r$. 
The function $\Delta$ has two roots: $r^{}_{\pm} \equiv M_{\MBH} \pm \sqrt{M_{\MBH}^2-a^2}$; 
the positive one $r^{}_{+}$ ($\geq r^{}_{-}$) coincides with the location of the event horizon.
The Kerr geometry is a stationary, axisymmetric and vacuum solution to the Einstein equations, 
which has has an additional symmetry described by a Killing tensor. 

In the limit where the SCO has zero mass ($m^{}_{\ast}\rightarrow 0$), 
it follows timelike geodesics of the MBH geometry which, in terms of proper time,
can be parameterized as $z^{\mu}(\tau) = (t(\tau),r(\tau),\theta(\tau),\phi(\tau))$.
Given the symmetries of the Kerr metric, each of these geodesics can be fully characterized by 
three constants of motion (per SCO unit mass): The energy (stationarity), $E$;
the angular momentum in the MBH spin direction (axisymmetry), $L^{}_{z}$, and the
{\em Carter constant} (Killing tensor symmetry), $Q$ (or $C \equiv Q -  
\left(L^{}_{z} - a E \right)^{2}$).  Using these symmetries we can completely separate the
geodesic equations so that they can be written in the form:
\ba
\rho^{2}\, \frac{d{t}}{d\tau} & = & \frac{1}{\Delta}\left( \Sigma^{2} E - 2 M_{\MBH} a r {L}^{}_{z} \right)\,,  
\label{tdot-GR} \\
\rho^{4}\, \left(\frac{d {r}}{d\tau} \right)^{2} & = &\left[ \left( r^{2} + a^{2} \right) E - a L^{}_{z} \right]^{2} - 
\left(Q+r^{2}\right)\Delta \,, \label{rdot-GR} \\
\rho^{4}\, \left(\frac{d{\theta}}{d \tau} \right)^{2} & = & C - \cot^{2}{\theta}{L}^{2}_{z} 
- a^{2} \cos^{2}{\theta} \left(1 - {E}^{2} \right)\,,\label{thetadot-GR} \\
\rho^{2}\, \frac{d \phi}{d{\tau}} & = & \frac{1}{\Delta}\left[ 2M_{\MBH}\,a\,r{E} + \frac{{L}^{}_{z}}{\sin^{2}\theta} 
\left( \Delta -a^{2}{\sin^{2}{\theta}} \right)  \right]\,, \label{phidot-GR}
\ea
where $\Sigma^{2} \equiv (r^2 + a^2)^{2} - a^{2}\Delta\,\sin^{2}\theta\,$.
For the purposes of EMRI modeling we restrict ourselves to bound geodesics (orbits), which can also be
characterized in terms of orbital elements: the eccentricity $e$, the
semilatus rectum $p$, and the inclination $\theta^{}_{\INC}$.  The first two are defined in terms of the
turning points (extrema) of the radial motion, the pericenter and apocenter ($r^{}_{\PERI}$ and $r^{}_{\APO}$),
by the Newtonian relations: $ r^{}_{\PERI} = pM^{}_{\bullet}/(1+e)$ and
$r^{}_{\APO} = pM^{}_{\bullet}/(1-e)\,.$ Likewise, the inclination angle $\theta^{}_{\INC}$ is
associated with the turning point of the polar motion, i.e. the minimum of $\theta$ 
($\theta^{}_{\MIN}\in [0,\pi/2]$), via the relation: 
$\theta^{}_{\INC} = {\rm sign}(L^{}_{z})\left(\pi/2 - \theta^{}_{\MIN}\right).$  Another useful
definition of the orbital inclination angle is: $\cos\iota = L^{}_{z}/\sqrt{L^{2}_{z}+C}\,.$

Alternatively, the orbits can also be characterized in
terms of three {\em fundamental} frequencies (see e.g.~\cite{Schmidt:2002qk,Drasco:2003ky,Fujita:2009bp}) with
respect to the BL time $t$ (they can be also constructed using proper time or any other 
time): 
$\Omega^{}_{r}$, associated with the radial motion (from periapsis to apoapsis and back); 
$\Omega^{}_{\theta}$, associated with polar motion; and 
$\Omega^{}_{\phi}$, associated with azimuthal motion.  Precessional 
orbital effects are due to mismatches between these frequencies, and they can be used to decompose, among other things, 
the GWs in a Fourier expansion~\cite{Drasco:2003ky}.  

%%%%%%%%%%%%%%%%%%%%%%%%%%%%%%%%%%%%%%%%%%%%%%%%%%%%%%%%%%%%%%%%%%%%%%
\subsection{Chimera Osculating Trajectories}
The foundations of the gravitational backreaction of a moving massive body on its own trajectory were
laid down in the papers by Mino, Tanaka, and Sasaki~\cite{Mino:1997nk} and Quinn and Wald~\cite{Quinn:1997am}.
These works established that the trajectory of a point mass can be understood as geodesic motion
in the background geometry modified by a local force, the {\em self-force}, given in terms
of the gradients of the {\em regularized}\footnote{For a point mass, the first-order metric perturbation is singular at the
particle location, and a regularization prescription~\cite{Detweiler:2002mi} must be employed to separate out the piece
responsible of the gravitational backreaction.} first-order metric perturbations.
Then, the equation of motion for the SCO, the MiSaTaQuWa equation, is 
\be
\frac{d^{2}z^{\alpha}}{d \tau^{2}}+
{\Gamma}^{\alpha}_{\mu \nu} u^{\mu} u^{\nu} = m_{\SCO}^{-1}
F_{\SF}^{\alpha}\,, \label{misataquwa1}
\ee
where the self-force, $F_{\SF}^{\alpha}$, is given by
\be
F_{\SF}^{\alpha} = - \frac{1}{2} m_{\SCO} \left({\met}^{\alpha \lambda} + u^{\alpha}u^{\lambda}\right)
u^{\mu}u^{\nu}
\left( 2 {\nabla}^{}_{\mu} h^{\REG}_{\nu\lambda} - {\nabla}^{}_{\lambda} 
h^{\REG}_{\mu \nu} \right)\,, \label{misataquwa2}
\ee
and $h^{\REG}_{\mu\nu}$ denotes the regularized metric perturbations.  This equation 
can be also reinterpreted as a geodesic equation for a point particle in the perturbed geometry 
$\met^{}_{\alpha\beta}+h^{\REG}_{\alpha\beta}$~\cite{Detweiler:2000gt}.
In order to evolve the SCO's trajectory subject to
the self-force in a self-consistent way, we use the method proposed in~\cite{Pound:2007th}, 
which is a relativistic extension of the \emph{osculating orbit method}.
The main idea is to model the SCO's worldline as a sequence of geodesics tangent to the worldline at any given point. 
The transition is smooth since EMRI trajectories remain close to a geodesic for a significant amount of time.
In principle, to carry out this transition we need to take into account all the parameters that 
characterize the {\em instantaneous} geodesic (the {\em principal} orbital elements) and the position of
the SCO within that geodesic (the {\em positional} orbital elements).  
In the initial version of the Chimera scheme~\cite{Sopuerta:2011te} we only consider the dissipative effects 
of the self-force, i.e.~those that only affect to principal orbital elements. Then, the evolution
of $(E,L^{}_{z},C/Q)$ is given by
\ba
&&\frac{dE}{d\tau} =  - {\zeta}^{(t)}_{\alpha} a^{\alpha}_{\SF}\,, \qquad
\frac{d{L}^{}_{z}}{d\tau} =  {\zeta}^{(\phi)}_{\alpha} {a}^{\alpha}_{\SF}\,, 
\label{evol-cons-motion-ELz}  \\
&&\frac{d{Q}}{d\tau} =  2\, {\xi}^{}_{\alpha \beta} {u}^{\alpha} a_{\SF}^{\beta}\,, \qquad
\frac{d{C}}{d\tau} = \frac{d{Q}}{d\tau} + 2 \left(a E -  L^{}_{z} \right) 
\left(\frac{d{L}^{}_{z}}{d\tau}  - a \frac{d{E}}{d\tau} \right) \,, \label{evol-cons-motion-QC}
\ea
where the SCO self-acceleration $a^{\alpha}_{\SF}$ is related to the self-force 
by $a^{\alpha}_{\SF} = m^{-1}_{\SCO} F^{\alpha}_{\SF}$, and where
${\zeta}^{(t)}_{\alpha}$ and ${\zeta}^{(\phi)}_{\alpha}$ are the timelike and axial
Killing vectors respectively, and ${\xi}^{}_{\alpha \beta}$ is the Killing tensor.  
%From the evolution of $(E,L^{}_{z},C/Q)$ we can find the evolution of $(e,p,\iota/\theta^{}_{\INC})$.

\subsection{Multipolar Post-Minkowskian Self-Acceleration}

Since the self-force is not known for generic Kerr orbits, we approximate it here via a multipolar, post-Minkowskian 
expansion (see e.g.~\cite{Blanchet:1984wm,Iyer:1993xi,Iyer:1995rn}). We begin by recasting the post-Minkowskian 
metric perturbation in the far field as the Kerr metric of the MBH $\met_{\mu \nu}^{\KERR,\HAR}$ in harmonic coordinates
(we shall discuss the issue of coordinates in next section), plus perturbations induced by the SCO. The latter
can be decomposed into conservative ({\emph{time-symmetric}}) and dissipative (\emph{time-asymmetric}) perturbations $h^{\RR}_{\mu\nu}$. 
Neglecting the former, we then have
\ba
\met^{\PMM}_{\mu\nu} &=& \met_{\mu\nu}^{\KERR,\HAR} + h^{\RR}_{\mu\nu}\,,
\label{resummed-metric}
\ea
The metric perturbations $h^{\RR}_{\mu\nu}$ are the quantities we use to describe the \emph{radiation-reaction} effects, and hence 
they are our approximation to the regularized metric perturbations $h^{\REG}_{\alpha\beta}$
of Eqs.~\eqref{misataquwa1} and~\eqref{misataquwa2}. The rate of change of the principal orbital elements in then described by 
Eq.~\eqref{evol-cons-motion-ELz}, with the force of Eq.~\eqref{misataquwa2} and the regularized metric perturbation of Eq.~\eqref{resummed-metric}.

The regular metric perturbation is a function of certain time-asymmetric scalar,
$V^{}_{\RR}$, and vector, $V_{\RR}^{i}$, radiation-reaction potentials, which to lowest order can be written as 
\be
h_{tt}^{\RR} = 2 \,V^{}_{\RR}\,, \qquad
h_{ti}^{\RR} = - 4 \,V^{i}_{\RR}\,, \qquad
h_{ij}^{\RR} = 2\, V^{}_{\RR}\,\delta_{ij} \,, \label{metricRR}
\ee
and the expressions for these potentials themselves are~\cite{Blanchet:1984wm,Blanchet:1996vx} 
\ba
V^{}_{\RR}(t^{}_{\HAR},\bm{x}^{}_{\HAR}) &=&  - \frac{1}{5} x^{ij}_{\HAR} M_{ij}^{(5)}(t_{\HAR}) 
+  \frac{1}{189} x^{ijk}_{\HAR} M_{ijk}^{(7)}(t_{\HAR}) 
 -  \frac{1}{70} \bm{x}^{2}_{\HAR} x^{ij}_{\HAR} M_{ij}^{(7)}(t_{\HAR}) + {\cal{O}}(x_{ij}^{\HAR} M_{ij}^{(9)})\,, 
\label{scalar_rr_potential} \\
V^{i}_{\RR}(t^{}_{\HAR},\bm{x}^{}_{\HAR})&=& \frac{1}{21} {x}^{<ijk>}_{\HAR} M_{jk}^{(6)}(t_{\HAR}) 
 -  \frac{4}{45} \epsilon_{ijk} x^{jl}_{\HAR} S_{kl}^{(5)}(t_{\HAR}) + {\cal{O}}(x^{ijk}_{\HAR} M_{ij}^{(8)})\,,
\label{vector_rr_potential}
\ea
where $(x^{\alpha}_{\HAR}) = (t^{}_{\HAR},x^i_{\HAR})$ are spacetime harmonic coordinates, 
$\epsilon^{}_{ijk}$ is the antisymmetric Levi-Civita symbol, and 
\be
\hat{x}^{<ijk>}_{\HAR} \equiv x^{ijk}_{\HAR} - \frac{3}{5}\bm{x}^2_{\HAR} \delta^{(ij}_{}x^{k)}_{\HAR}\,.
\ee 
is the symmetric trace-free (STF) projection of the multi-index quantity $x^{ijk} = x^{i} x^{j} x^{k}$.
The quantities $M_{ij}^{(n)}$, $M_{ijk}^{(n)}$, and $S_{ij}^{(n)}$ are the $n$th-time-derivative of the 
STF mass quadrupole, mass octopole and current quadrupole multipole moments of the source. 
The first term in $V^{}_{\RR}$ [Eq.~\eqref{scalar_rr_potential}] 
corresponds to the well-known Burke-Thorne radiation reaction potential~\cite{Burke:1970wx}.

The Chimera local self-force is completely specified once we prescribe how the multipole moments
depend on the orbital trajectories. This can be achieved by asymptotically matching a PN and a post-Minkowskian
solution in some buffer zone where both approximations are valid. The current Chimera implementation employs 
the leading-order expressions for these multipoles. As such, we are not consistently keeping a given PN order, 
but instead using leading-order expressions for the multipole moments, while keeping several such moments 
in the expansions.  To lowest order, the mass moments are given by 
\be
M^{}_{ij} = \eta\, m \; z^{}_{<ij>}\,,
\qquad
M^{}_{ijk} = \eta\, \delta m  \; z^{}_{<ijk>}
\label{mass-moments}
\ee
and the current quadrupole moment is
\be
S_{ij} = \eta\,\delta m\; \epsilon^{}_{kl <i} z_{j>}{}^{k} \dot{z}^{l}\,,
\label{current-moments}
\ee
where $\eta = m^{}_{\SCO}M^{}_{\MBH}/(m^{}_{\SCO}+M^{}_{\MBH})^{2}$ is the symmetric mass ratio 
and $\delta m = M^{}_{\MBH} - m^{}_{\SCO}$ is the mass difference.
To higher order, these moments become more complicated 
as there are now non-linear contributions from the non-reactive potentials (i.e.~non-linear 
contributions from the background) as well as tail and memory contributions.

\subsection{From Boyer-Lindquist to Harmonic Coordinates}
\label{coord-sec}

From the previous discussion it is clear that the MBH metric 
must be written in harmonic coordinates in order for all terms to be in the same coordinate
system and to be consistent with other ingredients of the Chimera scheme, in particular the
waveform generation procedure that we describe in next section. Nevertheless, there are parts of the
scheme that are easier to implement in BL coordinates, such as the integration of the
geodesic equations.  Therefore, it is crucial to find a map from BL to harmonic 
coordinates. 

Harmonic coordinates refer to coordinate systems, $\{x^{\alpha}_{\HAR}\}$, 
that satisfy $\square\, x^{\alpha}_{\HAR} = 0\,,$ 
where $\square$ is the D'Alembertian operator: 
$\square\equiv \met^{\alpha\beta}\nabla^{}_{\alpha}\nabla^{}_{\beta}$.  
In~\cite{1987PThPh..78.1186A}, based on 
work by Ding (see Ref.~\cite{1987PThPh..78.1186A} for references on this), a transformation 
to harmonic coordinates was found. 
Following~\cite{1987PThPh..78.1186A}, we can map the Kerr metric from BL 
coordinates $(t,r,\theta,\phi)$ to harmonic  ones $(t^{}_{\HAR},x^{}_{\HAR},y^{}_{\HAR},z^{}_{\HAR})$,
and the converse, via the coordinate transformations
\be
\begin{array}{ll}
t^{}_{\HAR} = t\,, &  t = t^{}_{\HAR}\,, \\
x^{}_{\HAR} = \sqrt{\left(r - M_{\MBH}\right)^{2} + a^{2}} \; \sin{\theta} \cos[\phi - \Phi(r)]\,, &
r = M_{\MBH} + \left[ \frac{r_{\HAR}^2 - a^2}{2} + \sqrt{\left(\frac{r_{\HAR}^2 - a^2}{2}\right)^{2} 
+ a^2 z_{\HAR}^2 }\right]^{1/2}\,,  \\  
y^{}_{\HAR} = \sqrt{\left(r - M_{\MBH}\right)^{2} + a^{2}} \; \sin{\theta} \sin[\phi - \Phi(r)]\,, &
\theta = \arccos\left(\frac{z^{}_{\HAR}}{r-M_{\MBH}}\right)\,, \\
z^{}_{\HAR} = \left(r - M_{\MBH}\right) \; \cos{\theta}\,, & 
\phi = \Phi(r) + \arctan\left(\frac{y^{}_{\HAR}}{x^{}_{\HAR}}\right)\,,
\end{array} \label{BL-to-H-to-BL}
\ee
where $r_{\HAR}\equiv(x_{\HAR}^{2} + y_{\HAR}^{2} + z_{\HAR}^{2})^{1/2}$
and the angle function $\Phi(r)$ is given by 
\ba
\Phi(r) = \frac{\pi}{2} - \arctan\left\{\frac{\frac{r-M_{\MBH}}{a} + 
\Omega(r)}{1 - \frac{r-M_{\MBH}}{a}\,\Omega(r)}\right\}\,, ~~ \mbox{with}~~
\Omega(r)  = \tan\left[\frac{a}{2\sqrt{M_{\MBH}^2-a^2}}\ln
\left(\frac{r-r^{}_{-}}{r-r^{}_{+}}\right)\right]\,. \label{omega-of-r}
\ea
%

%---------------------------------------------
\subsection{Waveform Generation}\label{waveform-generation}
Given the orbital evolution in harmonic coordinates one can construct the gravitational waveforms
through multipolar, post-Minkowskian expressions in terms of a sum over multipole moments
(see e.g.~\cite{Blanchet:2002av} for a review). This prescription is fully specified once expressions for the radiative moments are given in 
terms of derivatives of the orbital trajectory. Such identification, however, is difficult, 
as these moments are defined in the {\emph{far-zone}}  and have no knowledge of the {\emph{source}} multipole 
moments, defined in the {\emph{near zone}}.
In this initial version of the Chimera scheme, we only consider the leading-order contributions 
to the multipoles moments, i.e.~we identify the source and radiative moments. Then, in a transverse-traceless gauge 
and keeping multipole moments that include both the mass hexadecapole 
and the current octopole (thus keeping contributions one order higher than traditional kludge waveforms),
we have:
\ba
h_{ij}^{\TT} = \frac{2}{r} \ddot{M}_{ij}^{\STF} + \frac{2}{3 r} \left[ \dddot{M}_{ijk} n^{k} 
+ 4 \epsilon^{kl}{}_{(i} \ddot{S}_{j)k} n_{l}\right]^{\STF}  + 
\frac{1}{6r} \left[ \ddddot{M}_{ijkl} n^{k} n^{l}
+ 6  \epsilon^{kl}{}_{(i} \dddot{S}_{j)km} n^{l} n^{m}\right]^{\STF}\,.
\label{hTT}
\ea
The expressions for these multipole moments has been given in Eqs.~\eqref{mass-moments} and~\eqref{current-moments},
except for the mass hexadecapole and current octopole multipoles, which are given respectively by
\be
M^{}_{ijkl} = \eta\,m\;z^{}_{<ijkl>}\,, \qquad
S^{}_{ijk} = \eta\,m\;\epsilon^{}_{lm<i}z^{}_{jk>}{}^{l}\,\dot{z}^{m}\,. \label{current-octopole}
\ee
The plus and cross-polarized projections can then be constructed via 
\be
h^{}_{+,\times} = e^{ij}_{+,\times} h_{ij}^{\TT},
\label{h+x}
\ee
where $e^{ij}_{+,\times}$ are the plus- and cross-polarization tensors.
Finally, the observables, i.e.~the GW response function, are 
given by a projection of the plus- and cross-polarized waveform with the beam pattern functions 
of the detector.  For a detector like LISA, there are two such functions, $F_{+,I/II}$ and $F_{\times,I/II}$ 
(see, e.g.~\cite{Cutler:1998cc,Barack:2003fp}) and the response
is thus
\be
h \equiv \frac{\sqrt{3}}{2\,D^{}_{L}} \sum_{A=I,II}^{}\left(F^{}_{+,A}\, h^{}_{+} 
+ F^{}_{\times,A} h^{}_{\times}\right) \,,
\label{response}
\ee
where $D^{}_{L}$ is the luminosity distance from the source to the observer and the prefactor of 
$\sqrt{3}/2$ is due to the triangular arrangement of the LISA spacecraft constellation.

%%%%%%%%%%%%%%%%%%%%%%%%%%%%%%%%%%%%%%
\section{Numerical Implementation and some Numerical Results}
\label{num-implementation}

One of the main ingredients of the scheme is the integration of the 
geodesic equations~\eqref{tdot-GR}-\eqref{phidot-GR}.  The resulting equations for
the radial and polar BL coordinates, Eqs.~\eqref{rdot-GR} and~\eqref{thetadot-GR} respectively, 
have turning points (at pericenter and apocenter in the case of the radial coordinate, and at
the location of the orbital inclination angle in the case of the polar coordinate).  At these
points we have either $\dot{r} = 0$ or $\dot{\theta} = 0$ and the ODE solvers may encounter convergence
problems. To avoid these problems, we use angle variables associated with radial and polar
BL coordinates via 
\be
r = \frac{p M_{\MBH}}{1 + e \cos{\psi}}\,, 
\qquad 
\cos^{2}{\theta} = \cos^{2}{\theta}_{\rm min} \cos^{2}{\chi}\,. \label{ODEanglestoBL}
\ee
For convenience, we parametrize the trajectory in terms of the BL coordinate time $t$, which is
also a time harmonic coordinate, instead of the proper time $\tau$.  Then, the direct result
of the ODE integration is the sequence $(t,\psi(t),\chi(t),\phi(t))$ for $t\in [t^{}_{\rm ini},
t^{}_{\rm end}]$.

The main challenge in the numerical implementation of the Chimera scheme is the
evaluation of the time derivatives of the different quantities involved.  To understand 
this problem let us focus on the computation of our multipolar, post-Minkowskian self-force.
The radiation reaction potentials $V^{}_{\RR}$ and $V^{i}_{\RR}$ in~\eqref{scalar_rr_potential} and~\eqref{vector_rr_potential}
depend on up to the seventh time derivative of the mass quadrupole
and octopole moments and up to the fifth time derivative of the current quadrupole moment.
But since one also needs certain time derivatives of these potentials [see Eq.~\eqref{misataquwa2}]
one must compute up to the eighth time derivative of the mass quadrupole and octopole moments and 
the sixth time derivative of the current quadrupole moment.  

In principle, the computation of these derivatives can be done 
analytically by using repeatedly the equations of motion, Eqs.~\eqref{tdot-GR}-\eqref{phidot-GR}.
The problem is that we need the time derivatives of the trajectory in harmonic coordinates, and to 
pass from the ODE angles $(\psi(t),\chi(t),\phi(t))$ to harmonic coordinates we first need to use 
Eq.~\eqref{ODEanglestoBL} to go from these angles to BL coordinates, and then 
Eqs.~\eqref{BL-to-H-to-BL} to go from BL to harmonic coordinates. 
Therefore, we would need to obtain analytically higher time-derivatives of the ODE angles (up to eighth order), 
which involves using Christoffel symbols of the Kerr metric and several of their derivatives, and also
to differentiate several times Eqs.~\eqref{ODEanglestoBL} and~\eqref{BL-to-H-to-BL}.
In practice, this makes the analytical computations unfeasible, even using modern computer 
algebra systems.  

Therefore, one needs to resort to numerical evaluations of these derivatives. First, notice that 
the trajectory, the velocity, and the acceleration can be computed
with high accuracy directly from the integration of the ODEs in Eqs.~\eqref{tdot-GR}-\eqref{phidot-GR}.
Then, using analytical expressions, we can directly obtain the
second time derivatives of the mass quadrupole and octopole moments and the first time derivative of
the current quadrupole moment.  Thus, we just need to compute numerically up to six additional time derivative of the
mass moments (i.e. starting from their second time derivative) and up to five additional time derivative
of the current quadrupole moment (i.e. starting from its first time derivative).

Computing numerical derivatives is a subtle task, which we tried to implement in different ways.  For instance, we experimented with 
many finite difference rules (involving from a few to more than 20 evaluation points), as well as other 
generic numerical differentiation techniques (such as numerical interpolation or Chebyshev differentiation).
We found that, in order to obtain decent performance, a substantial
amount of fine-tuning was necessary. To overcome this difficulty, we looked
for a method better adapted to our problem.  Notice that 
multipole moments are functionals of the trajectories, which are piecewise Kerr bounded geodesic, and 
bounded geodesics can be characterized by three fundamental frequencies (in the generic case).  Then, 
an arbitrary functional of Kerr orbits, say $f[\psi,\chi,\phi](t)$, can
be expanded in a multiple Fourier series of these frequencies as~\cite{Drasco:2003ky}
\be
f[\psi,\chi,\phi](t) = \sum^{}_{k,m,n} f^{}_{k,m,n} e^{-i\,
(k\Omega^{}_{r}+m\Omega^{}_{\theta}+n\Omega^{}_{\phi})\,t}\,, \label{multifourier}
\ee
where $(k,m,n)$ are integers running from $-\infty$ to $+\infty$ and $f^{}_{k,m,n}$
are complex coefficients such that $f^{}_{-k,-m,-n} = \bar{f}^{}_{k,m,n}$.  There
are three special cases in which this expansion is simplified: (i) Circular equatorial
orbits;  (ii) Equatorial non-circular orbits; (iii) Circular non-equatorial orbits. In case (i), 
the Fourier series contains only a single frequency, the azimuthal one $\Omega^{}_{\phi}$. 
In case (ii) and (iii), there are two independent frequencies, $(\Omega^{}_{r},\Omega^{}_{\phi})$ 
and $(\Omega^{}_{\theta},\Omega^{}_{\phi})$ respectively.

By modeling the required multipole moments as an expansion of the form of Eq.~\eqref{multifourier}, 
using a standard least-square fitting algorithm, time derivatives are then computed via
\be
f^{(N)}[\psi,\chi,\phi](t) =  \sum^{}_{k,m,n}{f}^{N}_{k,m,n} 
 e^{-i\,(k\Omega^{}_{r}+m\Omega^{}_{\theta}+n\Omega^{}_{\phi})\,t}\,, \label{multifourierderivative}
\ee
where ${f}^{N}_{k,m,n} = (-i)^{N}(k\Omega^{}_{r}+m\Omega^{}_{\theta}+n\Omega^{}_{\phi})^{N}\,f^{}_{k,m,n} \,.$
We tested this technique with different types of functions and we have
found that it is very robust and provides very high accuracy even for the highest
derivatives.  For instance, the sixth time derivative is typically accurate to 
one part in $10^{5}$, which is more than enough for our purposes.  

\begin{figure*}[htb]
\centering
\includegraphics[width=0.328\textwidth]{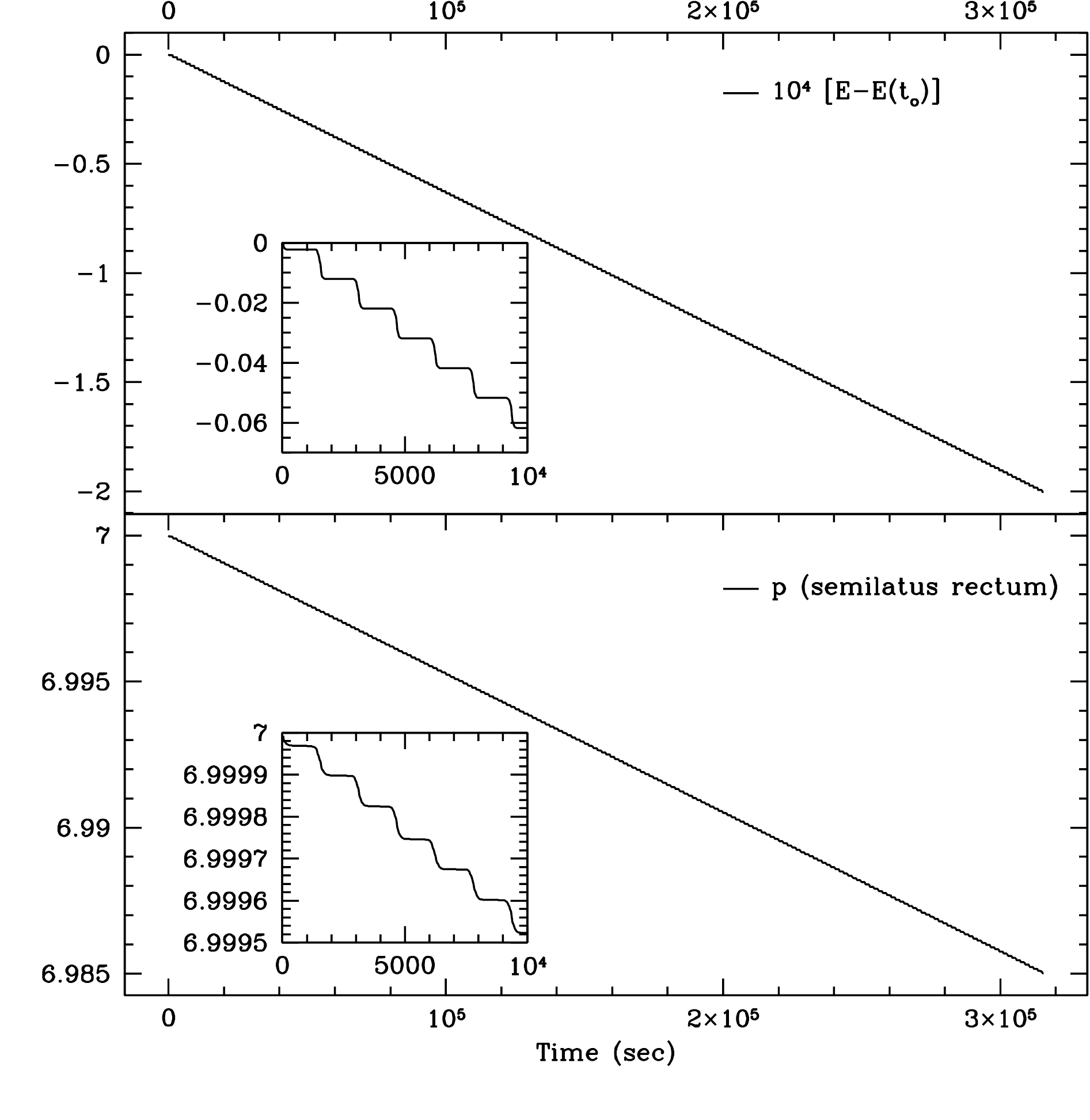}
\includegraphics[width=0.328\textwidth]{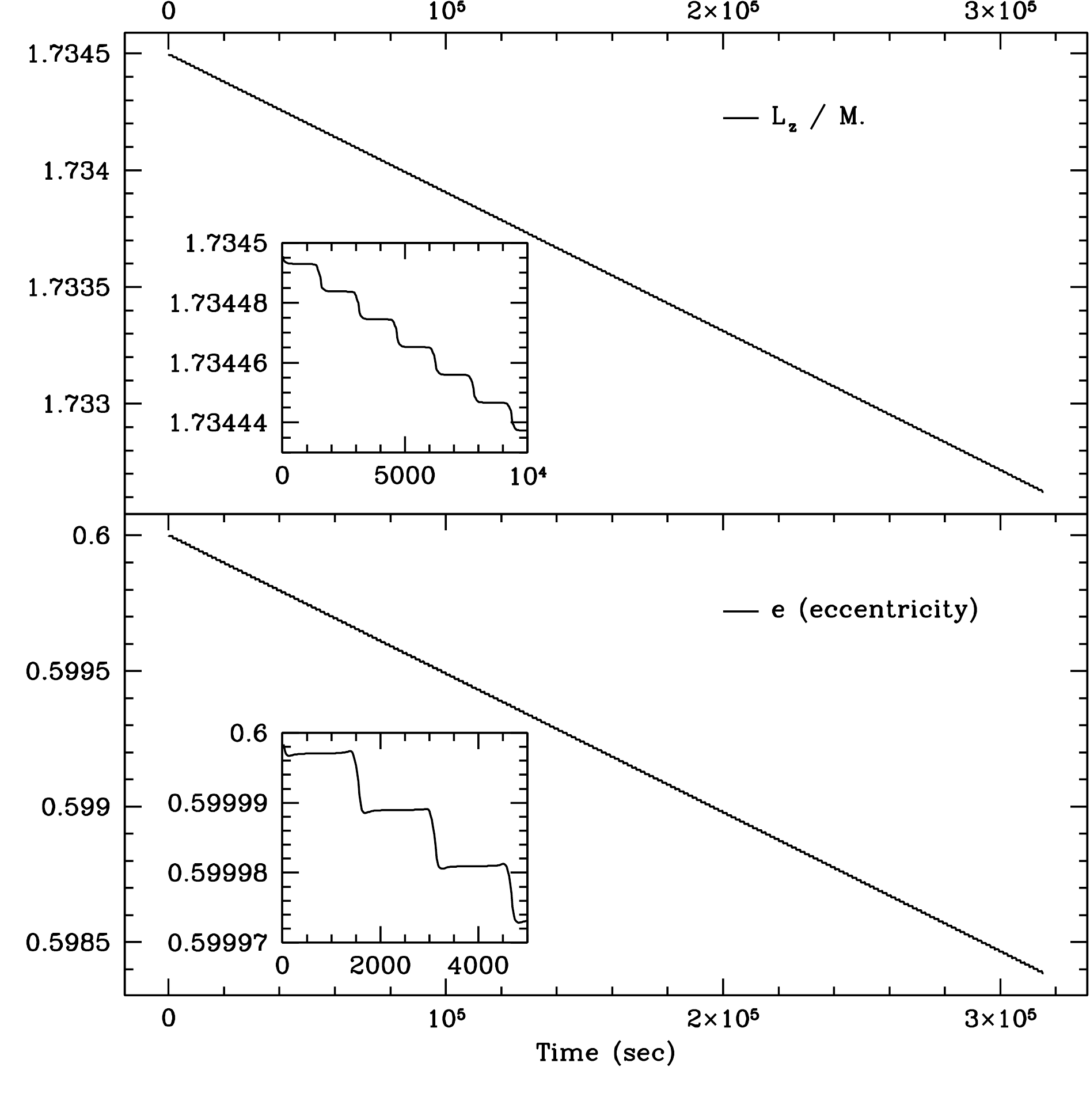}
\includegraphics[width=0.328\textwidth]{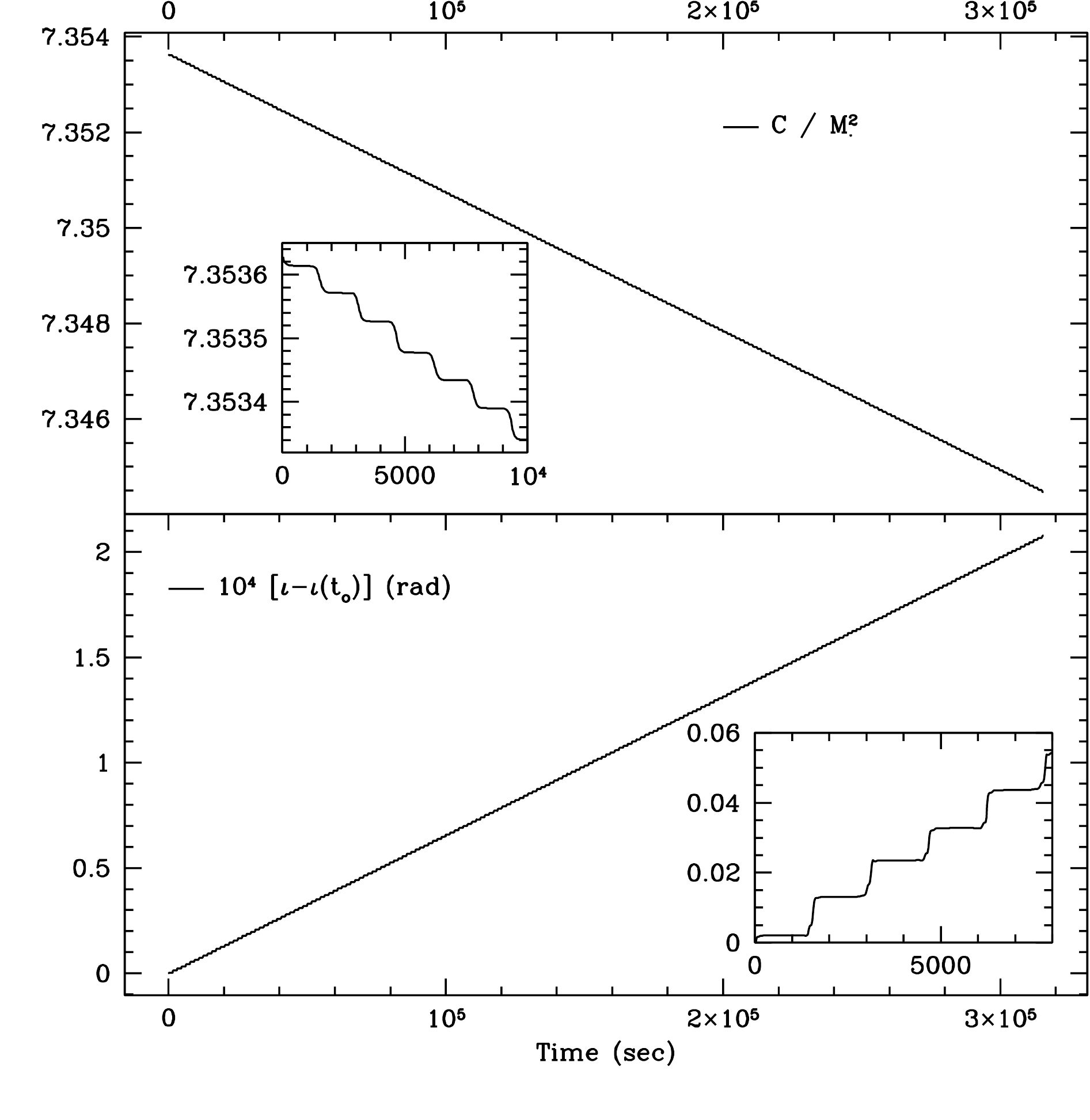}
\caption{Evolution (for a total time of $10^{-2}$ yrs) of an eccentric and inclined inspiral 
of a system characterized by: $M^{}_{\MBH} = 10^{6} M^{}_{\odot}$, $a/M^{}_{\MBH}=0.98$, 
and $q=10^{-5}$.  The plots show the evolution of the following quantities: Energy $E$ 
(top left), actually $10^{4}[E-E(t^{}_{o})]$, where $E(t^{}_{o}) = 0.9575513$; angular
momentum along the spin axis, $L^{}_{z}$ (top centre); Carter constant $C$ (top right);
semi-latus rectum, $p$ (bottom left), with $p(t^{}_{o}) = 7$; eccentricity, $e$ (bottom
centre), with $e(t^{}_{o}) = 0.6$; and inclination angle $\iota$ (bottom right), actually
$10^{4}[\iota-\iota(t^{}_{o})]$, where $\iota(t^{}_{o})=57.39$ deg.  All plots contain
subplots where the detailed evolution during a few orbital periods is
shown. \label{evolution-parameters-generic-orbit}}
\end{figure*}

In order to illustrate the strengths of the Chimera scheme, we present results 
for a generic, eccentric and inclined orbit.  The evolution of all the constants of 
motion/orbital elements for many orbital periods is shown in Figure.~\ref{evolution-parameters-generic-orbit}.
While the large scale evolution looks very smooth, the shorter scale evolution in the subplots
shows patterns with periods that match the orbital ones.  These patterns are a consequence of using a 
local in time self-force and do not appear in evolutionary schemes based on flux averaging over a certain number of
orbital periods. We can also appreciate from the global evolution that all quantities decay in time except for
the inclination, which grows. However, if we look at the evolution over
a few orbital periods, e.g. for the eccentricity, we can see that it can
grow locally in time although the global tendency is to decay. The Chimera
scheme leads to rich evolutionary patterns due to its local-in-time character, which
also makes it a very valuable tool to investigate questions like  
transient resonances~\cite{Flanagan:2010cd}.
%
%------------------------------------------------------------------------------%
%------------------------------------------------------------------------------%
%

%%%%%%%%%%%%%%%%%%%%%%%%%%%%%%%%%%%%%%
\section{Conclusions and Discussion} \label{discussion-conclusions}
We have presented the main ingredients of the Chimera scheme, a framework initially designed for the modeling
of EMRI dynamics and GW emission, but in principle also adaptable to  intermediate-mass ratio systems.  This scheme combines ingredients from 
the multipolar, post-Minkowskian formalism and black hole perturbation theory to evolve a non-geodesic wordline (with
respect to the MBH geometry) and to construct waveforms. The orbits are built as a sequence of local geodesics, 
whose orbital elements evolve according to a local self-force that we approximate via a multipolar, post-Minkowskian expansion. 
The leading-order term of the self-force corresponds to the well-known Burke-Thorne
radiation-reaction potential.  A crucial ingredient in this construction
is the mapping from BL coordinates, which we use to locally integrate the orbits, to harmonic coordinates, 
required both by the multipolar, post-Minkowskian 
self-force and by the GW multipolar expansion. Once we have trajectories in harmonic coordinates, 
it is straightforward to build waveforms.

The Chimera scheme can be improved in different ways. One possibility is to use more accurate, beyond leading-order, expressions
for the different multipole moments. Although these PN corrections are in general not known for
generic spinning binaries, they should certainly improve the accuracy of the Chimera evolutions for certain orbits. 
Another way to improve the scheme would
be to introduce conservative corrections to the background, for instance as it is done in the EOB
formalism~\cite{Buonanno:1998gg,Buonanno:2000ef,Damour:2009sm}.

A more detailed and exhaustive validation of the Chimera scheme would include, as a first step, a comparison of the
evolution of the constants of motion with those obtained by solving the Teukolsky equation~\cite{Hughes:1999bq,Hughes:2001jr}. 
One has to be careful in doing this comparison given 
that the latter employs averages over several cycles, while the Chimera fluxes are 
computed locally at the SCO's location. Once the fluxes have been validated, 
one should compare the waveforms themselves. An overlap study would determine 
the level of agreement between them.   

The Chimera scheme might also be useful to model IMRIs and even systems with moderate mass ratios. 
Recent comparisons between self-force, PN, and numerical relativity computations
have concluded that by replacing the mass ratio $q$ with  the symmetric
mass ratio $\eta$, the self-force predictions compare quite well with numerical 
relativity and PN predictions in the comparable-mass range~\cite{Barack:2010ny,LeTiec:2011bk}.
These results may allow for a simpler description of IMRIs, which otherwise would require 
either long, full numerical computations or higher-order perturbative computations, 
or a combination of both. In the Chimera scheme, $q\rightarrow \eta$ is actually
mostly built in already, since the mass ratio information enters through the definition
of the multipole moments. For these, we already use general binary expressions, 
instead of effective one-body ones with mass $q M^{}_{\MBH}$.

The ability to compute approximate local quantities at the location of the SCO,
in particular the multipolar, post-Minkowskian self-force,  
makes the Chimera scheme an interesting tool to study certain local behavior believed
to exist in EMRIs. For example, Flanagan and Hinderer~\cite{Flanagan:2010cd} have recently 
reported that certain rapid changes in orbital elements can arise for generic EMRIs when the 
orbital frequencies become commensurate. During these rapid changes, it has been 
suggested that EMRI waveforms may present {\em glitches}.  Reliable quantitative predictions
for these glitches are needed to assess the importance of this effect for gravitational 
wave astronomy. In this sense, we could compare the effect of these glitches in waveforms as 
computed from an averaged-scheme (such as the Teukolsky one) and a local one 
(such as the Chimera scheme).

%%%%%%%%%%%%%%%%%%%%%%%%%%%%%%%%%%%%%%%%%%%%%%%%%%%%%%%%%%%%%%%%%%%%%%%%%%%%%%
\section*{Acknowledgments}
CFS acknowledges support from the Ram\'on y Cajal Programme of the Spanish Ministry 
of Education and Science, by a Marie Curie
International Reintegration Grant (MIRG-CT-2007-205005/PHY) within the 7th
European Community Framework Programme, and from contract AYA-2010-15709
of the Spanish Ministry of Science and Innovation. 
NY acknowledges support from NSF grant PHY-1114374, as well as support provided by the National Aeronautics and Space Administration through Einstein Postdoctoral Fellowship Award Number PF0-110080, issued by the Chandra X-ray Observatory Center, which is operated by the Smithsonian Astrophysical Observatory for and on behalf of the National Aeronautics Space Administration under contract NAS8-03060. NY also acknowledges support from NASA grant NNX11AI49G, under sub-award 00001944.
We acknowledge the computational resources 
provided by the  Barcelona Supercomputing Centre (AECT-2011-3-0007) and 
CESGA (CESGA-ICTS-$200$). 

%%%%%%%%%%%%%%%%%%%%%%%%%%%%%%%%%%%%%%%%%%%%%%%%%%%%%%%%%%%%%%%%%%%%%%%%%%%%%%
\section*{References}
%\bibliography{master}

\begin{thebibliography}{10}
\expandafter\ifx\csname url\endcsname\relax
  \def\url#1{{\tt #1}}\fi
\expandafter\ifx\csname urlprefix\endcsname\relax\def\urlprefix{URL }\fi
\providecommand{\eprint}[2][]{\url{#2}}

\bibitem{Danzmann:2003tv}
Danzmann K and Rudiger A 2003 {\em Class. Quant. Grav.\/} {\bf 20} S1--S9

\bibitem{Prince:2003aa}
{Prince} T 2003 {\em American Astronomical Society Meeting\/} {\bf 202} 3701

\bibitem{AmaroSeoane:2007aw}
Amaro-Seoane P {\em et~al.\/} 2007 {\em Class. Quant. Grav.\/} {\bf 24}
  R113--R169 (\textit{Preprint} \eprint{astro-ph/0703495})

\bibitem{Barack:2003fp}
Barack L and Cutler C 2004 {\em Phys. Rev.\/} {\bf 69} 082005
  (\textit{Preprint} \eprint{gr-qc/0310125})

\bibitem{Collins:2004ex}
Collins N~A and Hughes S~A 2004 {\em Phys. Rev.\/} {\bf D69} 124022
  (\textit{Preprint} \eprint{gr-qc/0402063})

\bibitem{Glampedakis:2005cf}
Glampedakis K and Babak S 2006 {\em Class. Quant. Grav.\/} {\bf 23} 4167--4188
  (\textit{Preprint} \eprint{gr-qc/0510057})

\bibitem{Barack:2006pq}
Barack L and Cutler C 2007 {\em Phys. Rev.\/} {\bf D75} 042003
  (\textit{Preprint} \eprint{gr-qc/0612029})

\bibitem{Vigeland:2009pr}
Vigeland S~J and Hughes S~A 2010 {\em Phys. Rev.\/} {\bf D81} 024030
  (\textit{Preprint} \eprint{0911.1756})

\bibitem{Hughes:2010xf}
Hughes S~A 2010  (\textit{Preprint} \eprint{1002.2591})

\bibitem{Sopuerta:2010zy}
Sopuerta C~F 2010 {\em Gravitational Wave Notes\/} {\bf 4}(4) 3--47
  (\textit{Preprint} \eprint{1009.1402})

\bibitem{Hughes:2006pm}
Hughes S~A 2006 {\em AIP Conf. Proc.\/} {\bf 873} 233--240 (\textit{Preprint}
  \eprint{gr-qc/0608140})

\bibitem{Schutz:2009zz}
Schutz B~F 2009 {\em Class. Quant. Grav.\/} {\bf 26} 094020

\bibitem{Babak:2010ej}
Babak S, Gair J~R, Petiteau A and Sesana A 2011 {\em Class. Quant. Grav.\/}
  {\bf 28} 114001 (\textit{Preprint} \eprint{1011.2062})

\bibitem{Mino:1997nk}
Mino Y, Sasaki M and Tanaka T 1997 {\em Phys. Rev.\/} {\bf D55} 3457--3476
  (\textit{Preprint} \eprint{gr-qc/9606018})

\bibitem{Quinn:1997am}
Quinn T~C and Wald R~M 1997 {\em Phys. Rev.\/} {\bf D56} 3381--3394
  (\textit{Preprint} \eprint{gr-qc/9610053})

\bibitem{Barack:2009ey}
Barack L and Sago N 2009 {\em Phys. Rev. Lett.\/} {\bf 102} 191101
  (\textit{Preprint} \eprint{0902.0573})

\bibitem{Barack:2010tm}
Barack L and Sago N 2010 {\em Phys. Rev.\/} {\bf D81} 084021 (\textit{Preprint}
  \eprint{1002.2386})

\bibitem{Shah:2010bi}
Shah A~G, Keidl T~S, Friedman J~L, Kim D~H and Price L~R 2011 {\em Phys.
  Rev.\/} {\bf D83} 064018 (\textit{Preprint} \eprint{1009.4876})

\bibitem{Sopuerta:2011te}
Sopuerta C~F and Yunes N 2011 {\em Phys.Rev.\/} {\bf D84} 124060
  (\textit{Preprint} \eprint{1109.0572})

\bibitem{Blanchet:1984wm}
Blanchet L and Damour T 1984 {\em Phys. Lett.\/} {\bf A104} 82--86

\bibitem{Flanagan:2010cd}
Flanagan E~E and Hinderer T 2010  (\textit{Preprint} \eprint{1009.4923})

\bibitem{Schmidt:2002qk}
Schmidt W 2002 {\em Class. Quant. Grav.\/} {\bf 19} 2743 (\textit{Preprint}
  \eprint{gr-qc/0202090})

\bibitem{Drasco:2003ky}
Drasco S and Hughes S~A 2004 {\em Phys. Rev.\/} {\bf D69} 044015
  (\textit{Preprint} \eprint{astro-ph/0308479})

\bibitem{Fujita:2009bp}
Fujita R and Hikida W 2009 {\em Class. Quant. Grav.\/} {\bf 26} 135002
  (\textit{Preprint} \eprint{0906.1420})

\bibitem{Detweiler:2002mi}
Detweiler S and Whiting B~F 2003 {\em Phys. Rev.\/} {\bf D67} 024025
  (\textit{Preprint} \eprint{gr-qc/0202086})

\bibitem{Detweiler:2000gt}
Detweiler S 2001 {\em Phys. Rev. Lett.\/} {\bf 86} 1931--1934
  (\textit{Preprint} \eprint{gr-qc/0011039})

\bibitem{Pound:2007th}
Pound A and Poisson E 2008 {\em Phys. Rev.\/} {\bf D77} 044013
  (\textit{Preprint} \eprint{0708.3033})

\bibitem{Iyer:1993xi}
Iyer B~R and Will C~M 1993 {\em Phys. Rev. Lett.\/} {\bf 70} 113--116

\bibitem{Iyer:1995rn}
Iyer B~R and Will C~M 1995 {\em Phys. Rev.\/} {\bf D52} 6882--6893

\bibitem{Blanchet:1996vx}
Blanchet L 1997 {\em Phys. Rev.\/} {\bf D55} 714--732 (\textit{Preprint}
  \eprint{gr-qc/9609049})

\bibitem{Burke:1970wx}
Burke W~L 1971 {\em J. Math. Phys.\/} {\bf 12} 401--418

\bibitem{1987PThPh..78.1186A}
{Abe} M, {Ichinose} S and {Nakanishi} N 1987 {\em Progress of Theoretical
  Physics\/} {\bf 78} 1186--1201

\bibitem{Blanchet:2002av}
Blanchet L 2006 {\em Living Rev. Relativity\/} {\bf 9} 4 (\textit{Preprint}
  \eprint{gr-qc/0202016})

\bibitem{Cutler:1998cc}
Cutler C 1998 {\em Phys. Rev.\/} {\bf D57} 7089--7102

\bibitem{Buonanno:1998gg}
Buonanno A and Damour T 1999 {\em Phys. Rev.\/} {\bf D59} 084006
  (\textit{Preprint} \eprint{gr-qc/9811091})

\bibitem{Buonanno:2000ef}
Buonanno A and Damour T 2000 {\em Phys. Rev.\/} {\bf D62} 064015
  (\textit{Preprint} \eprint{gr-qc/0001013})

\bibitem{Damour:2009sm}
Damour T 2010 {\em Phys. Rev.\/} {\bf D81} 024017 (\textit{Preprint}
  \eprint{0910.5533})

\bibitem{Hughes:1999bq}
Hughes S~A 2000 {\em Phys. Rev.\/} {\bf D61} 084004

\bibitem{Hughes:2001jr}
Hughes S~A 2001 {\em Phys. Rev.\/} {\bf D64} 064004

\bibitem{Barack:2010ny}
Barack L, Damour T and Sago N 2010 {\em Phys. Rev.\/} {\bf D82} 084036
  (\textit{Preprint} \eprint{1008.0935})

\bibitem{LeTiec:2011bk}
Le~Tiec A, Mroue A~H, Barack L, Buonanno A, Pfeiffer H~P {\em et~al.\/} 2011
  {\em Phys. Rev. Lett.\/} {\bf 107} 141101 (\textit{Preprint}
  \eprint{1106.3278})

\end{thebibliography}

\providecommand{\newblock}{}

\end{document}